\documentclass{svproc}
\usepackage{graphicx}%
\usepackage{multirow}%
\usepackage{amsmath,amssymb,amsfonts}%
\usepackage{mathrsfs}%
\usepackage[title]{appendix}%
\usepackage{xcolor}%
\usepackage{textcomp}%
\usepackage{manyfoot}%
\usepackage{booktabs}%
\usepackage{algorithm}%
\usepackage{algorithmicx}%
\usepackage{algpseudocode}%
\usepackage{listings}%
\usepackage[normalem]{ulem}
\useunder{\uline}{\ul}{}
\usepackage{url}

\def\orcidID#1{\unskip$^{[#1]}$} 

\begin{document}
\mainmatter              
\title{Dense Subgraph Clustering and a New Cluster Ensemble Method}
\titlerunning{Dense Subgraph Clustering}  
%
\author{The-Anh Vu-Le\inst{1}\orcidID{0000-0002-4480-5535} 
\and Jo\~{a}o Alfredo Cardoso Lamy\inst{2}\orcidID{0009-0005-4744-4754} \and Tom\'as Alessi\inst{2}\orcidID{0009-0006-2658-5758} 
\and Ian Chen\inst{1}\orcidID{0009-0000-3777-3603}
\and Minhyuk Park\inst{1}\orcidID{0000-0002-8676-7565} 
\and Elfarouk Harb\inst{1}\orcidID{0000-0002-7770-0932} 
\and 
George Chacko\inst{1}\orcidID{0000-0002-2127-1892} \and Tandy Warnow\inst{1}\orcidID{0000-0001-7717-3514}
}
\authorrunning{Vu-Le et al.} 

\institute{$^1$School of Computing and Data Science, University of Illinois Urbana-Champaign, Urbana IL 61801\\
$^2$Insper Institute, Sao Paolo, Brazil\\
\email{warnow@illinois.edu},\\ WWW home page:
\texttt{https://tandy.cs.illinois.edu}}

\maketitle              

\begin{abstract}
We propose DSC-Flow-Iter, a new community detection algorithm that is based on iterative extraction of dense subgraphs. Although DSC-Flow-Iter leaves many nodes unclustered, it is competitive with leading methods and has high-precision and low-recall, making it complementary to modularity-based methods that typically have high recall but lower precision. Based on this observation, we introduce a novel cluster ensemble technique that combines DSC-Flow-Iter with modularity-based clustering, to provide improved accuracy.  We show that our proposed pipeline, which uses this ensemble technique, outperforms its individual components and improves upon the baseline techniques on a large collection of synthetic networks.
\end{abstract}


\section{Introduction}

Community detection, also known as graph clustering, is the task of partitioning a network's vertices into groups that exhibit strong community structure.  
In general, the groups should exhibit strong edge density (i.e., many edges connecting nodes within groups), they should be well-connected (i.e., they should not have small edge cuts), and  there should be relatively fewer edges between different groups \cite{yang2012defining}. 
Identifying these structures can be helpful for understanding the organization and function of complex systems across various domains, from bioinformatics, social networks, and scientific research networks
\cite{su2022comprehensive}. 

Community detection is often posed as an optimization problem, with modularity one of the best known optimization problems \cite{newman2004detecting}.
Since these problems are generally NP-hard, finding optimal solutions is not reliable, and heuristics, such as Leiden \cite{leiden} and Louvain \cite{louvain}, are used.  

The Iterative k-Core (IKC) algorithm \cite{ikc} is an example of an alternative approach, which focuses on the iterative extraction of high-quality subgraphs.
Given the network, IKC finds a $k$-core (i.e., a subgraph in which all nodes have degree at least $k$) for the largest achieved value of $k$,  treats the $k$-core as a community and then removes it, and recurses until a stopping rule applies (e.g., no $k$-cores for $k > 10$ are left). 

This iterative approach provides the motivation for our proposed method, which replaces the $k$-core concept with a different measure of density.
We examine clustering methods that use greedy iterative approaches based on seeking subgraphs that have the largest edge density \cite{harb2022faster}, where  the density of a subgraph is the ratio of the number of edges to the number of vertices.

The Dense Subgraph Decomposition procedure described in \cite{harb2022faster} partitions the vertex set of the input graph $G = (V,E)$ into a series of subsets based on decreasing density. 
The process begins by identifying $S_1$, which is the unique maximal densest subgraph in the graph $G$. 
Subsequently, for each step $i$, the procedure identifies the maximal set $S$ from the remaining vertices that maximizes the function $(|E(S)| + |E(S, U_{i-1})|) / |S|$, where $U_{i-1}$ is the union of all previously identified subsets $S_1, \dots, S_{i-1}$, $E(S)$ is the set of edges  between vertices of $S$ (i.e., $E(S) = \{ \{u, v\} \in E \mid u, v \in S \}$), and $E(S, S')$ is the  set of edges between vertices of $S$ and $S'$ (i.e., $E(S, S') = \{ \{u, v\} \in E \mid u \in S, v \in S'\}$). 
Notice that $|E(S)| / |S|$ is the density of the subgraph induced by $S$. Hence, this approach differs from a simple iterative extraction of the densest subgraph since it also accounts for the edges connecting a candidate subgraph to the parts that have already been extracted. 

Harb and colleagues  proposed two main approaches to compute a dense subgraph decomposition: 
FISTA \cite{harb2022faster}, which solves a convex relaxation of the problem using the Fast Iterative Shrinkage-Thresholding Algorithm (FISTA), and Flow \cite{harb2025corporate}, which solves a max-flow formulation of the problem using a push-relabel solver.
The Flow technique is expected to provide better accuracy than FISTA. 

We study the application of FISTA and Flow within an iterative framework, and find that our new approach (DSC-Flow-Iter), described in greater detail below, provides the best accuracy of these density-based methods. 
DSC-Flow-Iter is not quite as accurate as the best standard method we evaluated (i.e., Leiden-CPM, followed by postprocessing using WCC \cite{park2024improved-arxiv}), but a new cluster ensemble approach we present here that combines DSC-Flow-Iter with other techniques attains higher accuracy than all tested methods.

Due to space restrictions, full details and some results are provided in the Supplementary Materials, available at the DSC-Flow-Iter github site \cite{dsc-github}.


\section{Materials and Methods}
\label{sec:materials}

We present an overview of the methods we studied and the experiments we performed to evaluate these methods.
The data used in this study are available from prior publications (see \cite{ecsbm_benchmark}). 
The software for the DSC-Flow-Iter and ensemble methods are available at \cite{dsc-github} and \cite{clustermerger-github}.
The software for other methods as well as the specific commands used are detailed in Supplementary Materials Section 2. 

\subsection{Datasets}
\vspace{-.1in}
\paragraph{Real-world networks} 

We use a collection of $78$ real-world networks sourced from the Netzschleuder database \cite{netzschleuder} and from \cite{Park24-11-CM-journal}. These networks span various domains, including social networks, collaboration networks, web graphs, and communication networks. Their sizes range from $906$ to $14$ million of nodes. A complete list of these networks is available in the Supplementary Materials Section 1. For evaluating clustering accuracy of different methods, we excluded the $4$ largest networks (\texttt{livejournal}, \texttt{orkut}, \texttt{bitcoin}, and \texttt{CEN}) and generated synthetic networks based on the remaining $74$; the largest of these networks has $1.4$ million nodes. For evaluating computational performance, we used all $78$ networks. 

\vspace{-.1in}
\paragraph{Synthetic networks} 

We use EC-SBM \cite{vu2025ecsbm} to generate synthetic networks
based on  $74$ of the real-world networks we studied.
EC-SBM takes as input the parameters obtained from a clustered network and generates a synthetic network that aims to reproduce the cluster and network statistics. 
For this study, we used 74 of the real-world networks we studied, with the input clustering obtained using Stochastic Block Models (SBMs) computed using graph-tool \cite{graph-tool}, and then post-processed using Well Connected Clusters (WCC, see below) to ensure that the clusters are well-connected.
This input clustering approach was selected because 
 \cite{vu2025ecsbm} established that it provides the closest fit to the network and clustering statistics.

\subsection{Clustering Methods}

We compared density-based clustering methods to standard clustering methods, including
Leiden optimizing modularity (Leiden-Mod), Leiden optimizing under the Constant Potts Model (Leiden-CPM), Infomap, and IKC.
Of these, only the density-based methods are new, and so we describe these further.

\vspace{-.1in}
\paragraph{Density-based techniques}

We adapt the Flow and FISTA algorithms \cite{harb2022faster,harb2025corporate}  as follows.
Each of these methods computes values for the vertices so that all vertices with the same values will be put into the same cluster.
The values produced by FISTA are approximate and cannot be used naively.
Therefore, we apply a simple rounding heuristic, grouping nodes whose values round to the same integer. Since these groups may contain two or more connected components, we divide them into connected components, and call this method DSC-FISTA(int).
In contrast, the values produced by Flow can be used directly (i.e., no rounding is needed, although division into connected components is still appropriate), and we call this method DSC-Flow.

An alternative way to use the two algorithms is through the iterative extraction of the densest subgraph. That is, given an input network, we can compute its approximate densest subgraph and remove the subgraph as a community, then recurse on the remaining graph until all vertices are accounted for. For Flow, the approximate densest subgraph contains all the nodes with the highest vertex values. For FISTA, the fractional peeling technique proposed in \cite{harb2022faster} is used to get the approximate densest subgraph. Fractional peeling iteratively removes the vertex with the lowest current vertex value while updating the loads of its neighbors, and returns the densest subgraph generated during this process. When removing a subgraph, isolated vertices are produced; each such  vertex is   left as a singleton cluster. We name these methods DSC-Flow-Iter and DSC-FISTA-Iter, respectively.

\subsection{Post-processing techniques}

We utilized the post-processing techniques described in \cite{sbmwcc-conf,sbmwcc-journal-arxiv}, which are Connected Components (CC) and Well-Connected Clusters (WCC).
The CC method splits a cluster that is internally disconnected into multiple clusters, each containing one of its connected components; we always apply this to any method that produces internally disconnected components.
WCC is designed to refine an existing clustering by iteratively partitioning each cluster by removing minimum edge cuts until all the  clusters are well-connected. We use the threshold from \cite{Park24-11-CM-journal}, which is that the  size of its minimum edge-cut is greater than $\log_{10}(n)$ where $n$ is the number of nodes inside the cluster.  
 WCC is typically benign or beneficial when used with SBMs for community detection \cite{sbmwcc-conf,sbmwcc-journal-arxiv}. 
Here, we study the effect of WCC  as a post-processing technique on the clustering methods we study.

\subsection{The New Cluster Ensemble Technique}

Cluster ensembles (also known as consensus clustering methods) are designed to take in a collection of clustering methods and return a clustering that achieves better stability or accuracy.
There is a large literature on this problem, with recent surveys in \cite{zhang2022weighted,golalipour2021clustering,hussain2025parallel}, which we very briefly summarize here.
The Cluster Ensemble Selection problem addresses the question of, given the collection of clustering methods, which ones should be included in the ensemble?
Once the subset of clusterings is selected, different cluster ensembles use a variety of techniques to compute the consensus clustering.
A common technique is to compute the co-occurrence matrix, which indicates for each pair of nodes the fraction of the input clusterings in which the two nodes are in the same cluster.
This co-clustering matrix is then used to compute the consensus clustering.
Note that this type of approach inherently assumes that all the input clusterings place all (or nearly all) of the network nodes into clusters (equivalently, do not produce clusters of size 1).

In our study, we are interested in learning whether density-based clustering methods, which we find do not cluster a large fraction of the network, can be used within a cluster ensemble approach.
However, when some of the input clustering methods have very low node coverage,  the co-occurrence matrix may incorrectly identify pairs that are strongly supported as having low support.
In this case,  the ensemble methods are likely to have poor accuracy. For this reason, we designed a new cluster ensemble method that we now describe.

Given a network $N_0=(V,E)$, an arbitrary set of input clusterings, and a threshold value $t \geq 0$, we compute an ensemble clustering as follows.
First, we build a weighted network $N_1$, initialized to $N_0$. 
Some of the edges will be removed and the ones that remain will be weighted based on the following protocol.
For an edge $e = \{u,v\}$ in $N_0$ (and hence in $N_1$), we set $\tilde{M}_e$ to be the number of input clusterings that assign both $u$ and $v$ to the same cluster and $M_e$ is the number of input clusterings that assign $u$ and $v$ each to a non-singleton cluster. 
We  define the weight $w_e$ of an edge $e = \{v,e\} \in E$ as follows.
If no input clustering  clusters $u$ and $v$ together, then we set  $w_e=0$.  
For all other edges $e$, we set $w_e = \tilde{M}_e / M_e$; these edges all have positive weight.
The edge $e$ is deleted if $w_e < t$.

We then run a final community detection algorithm to cluster $N_1$. 
One approach is to cluster the weighted network $N_1$, which requires that the final clustering technique can handle weighted networks. 
Another approach is to cluster the unweighted version of $N_1$.


\subsection{Evaluation} 

To evaluate the methods, we compare them against the ground-truth partitions of our synthetic networks. We use three metrics: Adjusted Mutual Information (AMI) \cite{vinh2009information-AMI}, Adjusted Rand Index (ARI) \cite{hubert1985comparing}, and Normalized Mutual Information (NMI) \cite{forbes1995classification}. 
We also use False Positive Rate (FPR), precision, and recall, where we represent a clustering by the equivalence relation on pairs of nodes that it defines,  and we score an estimated clustering using this representation. 

\subsection{Experiments}

We conducted four experiments:
\begin{itemize}
    \item Experiment 1: We compare different density-based techniques and establish that DSC-Flow-Iter is the best of all the density-based techniques we studied.
    \item Experiment 2: We compare DSC-Flow-Iter with previous community detection methods.
    \item Experiment 3: Cluster ensemble design and evaluation. 
    \item Experiment 4: We benchmark the runtime of the proposed pipeline on real-world networks.
\end{itemize}

All experiments were conducted on the Illinois Campus Cluster, using nodes with 256GB of RAM. The computation time for each task (running a community detection algorithm on a network, or running the post-processing treatment on a clustering result) was limited to 3 days.

\section{Experimental results}
\label{sec:results}

\subsection{Experiment 1: Comparing density-based methods}

\begin{figure}[!ht]
    \centering
    \includegraphics[width=0.8\linewidth]{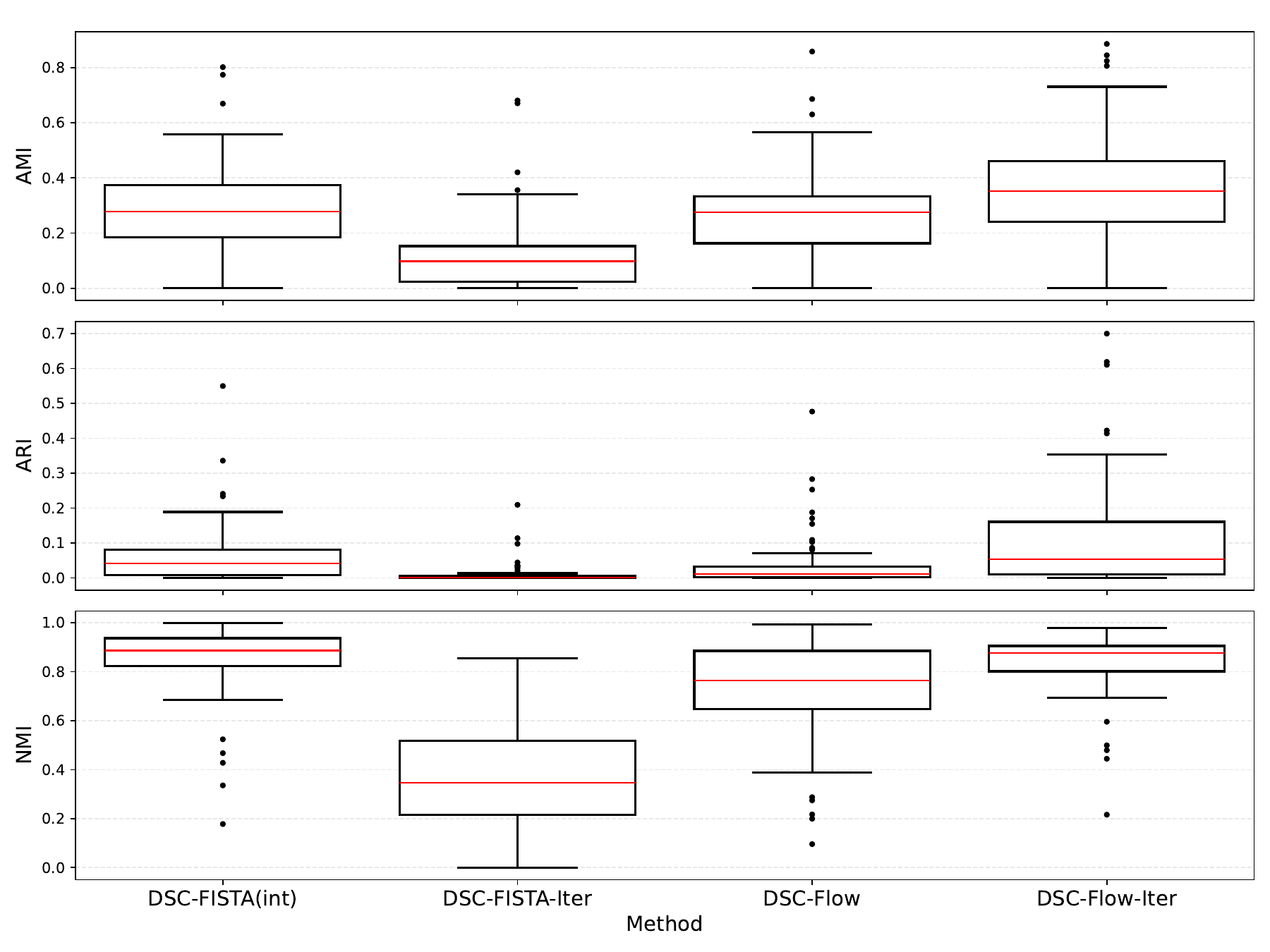}
    \caption[Comparison of density-based methods regarding community detection accuracy on synthetic networks]{\textbf{Accuracy of density-based clustering methods on synthetic networks} The results are on $74$ synthetic EC-SBM networks with SBM+WCC ground-truth clustering.  DSC-Flow-Iter is the method with the highest accuracy for all metrics.}
    \label{fig:exp1}
\end{figure}

We examine four different density-based methods: DSC-Flow, DSC-Flow-Iter,  DSC-FISTA(int), and DSC-FISTA-Iter. 
As shown in Figure \ref{fig:exp1}, DSC-Flow-Iter consistently achieves the highest accuracy across all three metrics. This result establishes that the iterative removal of the densest subgraph, as identified by the more precise Flow algorithm, is the most promising approach. Therefore, we select DSC-Flow-Iter as the representative method for our density-based approach in all subsequent experiments.
However, as noted in Supplementary Materials Table S1, DSC-Flow-Iter has low node coverage, on average only putting 63.5\% of the nodes into non-singleton clusters.

\begin{figure}[!ht]
    \centering
    \includegraphics[width=0.8\linewidth]{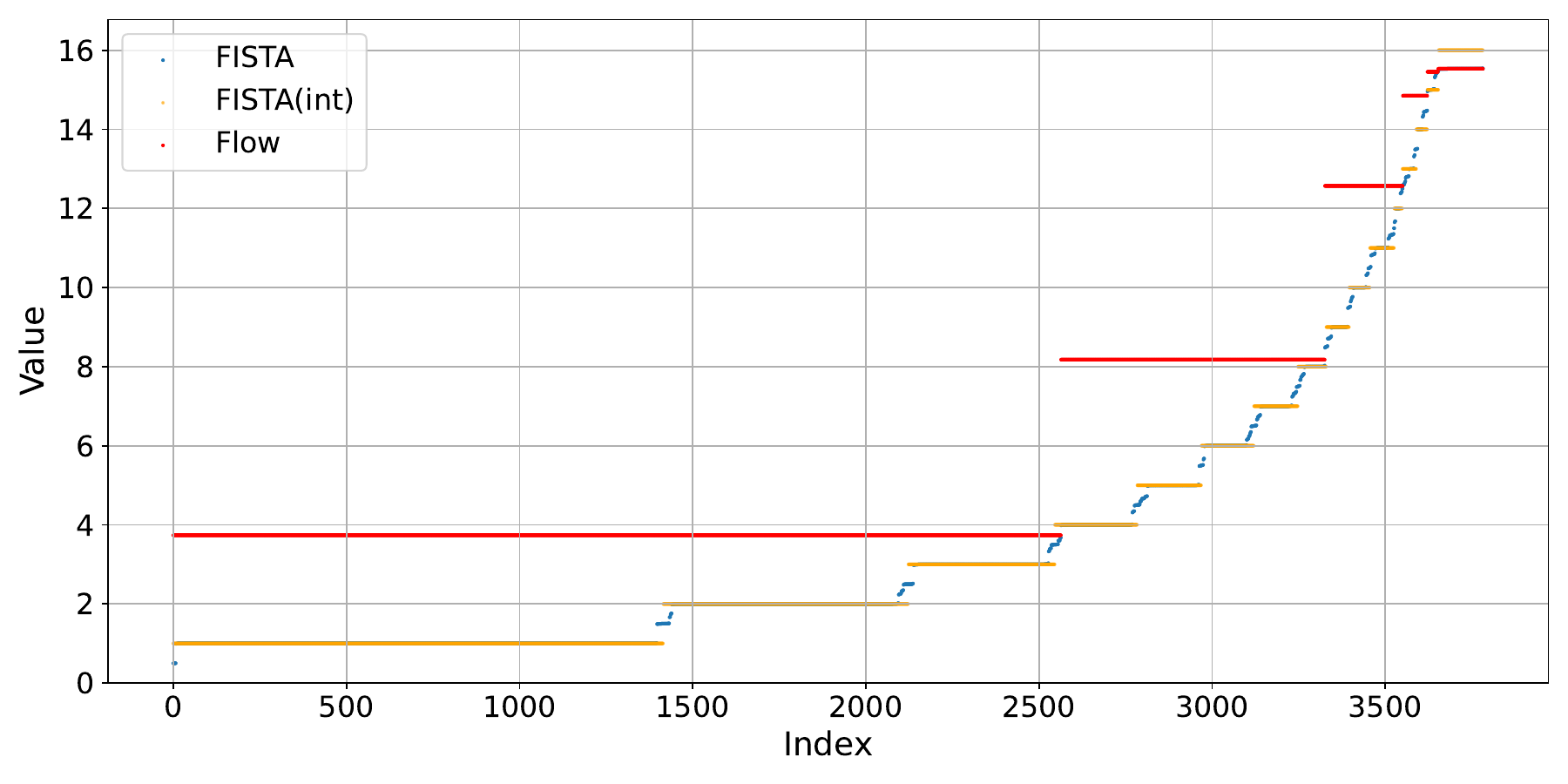}
    \caption[Vertex values output by density-based techniques]{\textbf{Vertex values output by density-based techniques} The result is on the EC-SBM synthetic network based on \texttt{bitcoin\_alpha} and SBM+WCC ground-truth clustering.}
    \label{fig:density/bitcoin_alpha}
\end{figure}

The performance differences between the methods can be understood by examining the vertex values they produce, as illustrated for a sample network in Figure \ref{fig:density/bitcoin_alpha}. The output of the FISTA algorithm forms a staircase-like pattern composed of many distinct steps connected by intermediate transitional values. This explains why integer rounding, although naive, is already an effective heuristic: the granular steps identified by FISTA are already well-separated, so the rounding process still preserves the underlying community structure. In contrast, the Flow algorithm produces a much coarser partitioning. As seen in the figure, a single, large step in the Flow output often encompasses and merges multiple, smaller steps from the FISTA output. For example, the first major plateau for Flow (at a value just under $4$) corresponds to a region where FISTA identifies several distinct, lower-level density steps (at values around $1$, $2$, and $3$). This shows that the Flow method effectively groups together many of the more granular communities suggested by FISTA into larger, more discrete clusters. 

\subsection{Experiment 2: Evaluating DSC-Flow-Iter}

We compared our best method, DSC-Flow-Iter, to Leiden-Mod, Leiden-CPM, Infomap, and IKC, all established community detection algorithms. 
The best parameter choices for IKC (the minimum value $k$ that is considered) and Leiden-CPM (the resolution parameter value) were determined in a preliminary experiment (Supplementary Materials Figure S1).

Table \ref{tab:exp2} shows the accuracy of these methods without post-processing using WCC.  Leiden-CPM(0.01) has the best AMI and ARI scores. Our proposed DSC-Flow-Iter is the next most accurate method for AMI and ARI accuracy, and the best for NMI. This demonstrates that DSC-Flow-Iter, although not the most accurate method, is competitive with other techniques. 


\begin{table}[]
\centering
\setlength{\tabcolsep}{4pt}
\renewcommand{\arraystretch}{1.5}
\caption[]{\textbf{Accuracy of community detection methods on synthetic networks} The results are on $74$ synthetic EC-SBM networks with SBM+WCC ground-truth clustering.
The best accuracy is boldfaced, the second-best accuracy is underlined.}
\label{tab:exp2}
\resizebox{\textwidth}{!}{%
\begin{tabular}{|l|lll|lll|lll|}
\hline
\multicolumn{1}{|c|}{\multirow{2}{*}{}} &
  \multicolumn{3}{c|}{\textbf{AMI}} &
  \multicolumn{3}{c|}{\textbf{ARI}} &
  \multicolumn{3}{c|}{\textbf{NMI}} \\ \cline{2-10} 
\multicolumn{1}{|c|}{} &
  \multicolumn{1}{c|}{\textbf{Med}} &
  \multicolumn{1}{c|}{\textbf{Avg}} &
  \multicolumn{1}{c|}{\textbf{Std}} &
  \multicolumn{1}{c|}{\textbf{Med}} &
  \multicolumn{1}{c|}{\textbf{Avg}} &
  \multicolumn{1}{c|}{\textbf{Std}} &
  \multicolumn{1}{c|}{\textbf{Med}} &
  \multicolumn{1}{c|}{\textbf{Avg}} &
  \multicolumn{1}{c|}{\textbf{Std}} \\ \hline
\textbf{DSC-Flow-Iter} &
  \multicolumn{1}{l|}{{\ul 0.352}} &
  \multicolumn{1}{l|}{{\ul 0.370}} &
  0.211 &
  \multicolumn{1}{l|}{{\ul 0.053}} &
  \multicolumn{1}{l|}{{\ul 0.121}} &
  0.157 &
  \multicolumn{1}{l|}{\textbf{0.876}} &
  \multicolumn{1}{l|}{\textbf{0.833}} &
  0.129 \\ \hline
\textbf{IKC(1)} &
  \multicolumn{1}{l|}{0.297} &
  \multicolumn{1}{l|}{0.307} &
  0.167 &
  \multicolumn{1}{l|}{0.019} &
  \multicolumn{1}{l|}{0.060} &
  0.104 &
  \multicolumn{1}{l|}{0.816} &
  \multicolumn{1}{l|}{0.802} &
  0.132 \\ \hline
\textbf{Leiden-CPM(0.01)} &
  \multicolumn{1}{l|}{\textbf{0.463}} &
  \multicolumn{1}{l|}{\textbf{0.435}} &
  0.273 &
  \multicolumn{1}{l|}{\textbf{0.126}} &
  \multicolumn{1}{l|}{\textbf{0.238}} &
  0.250 &
  \multicolumn{1}{l|}{{\ul 0.854}} &
  \multicolumn{1}{l|}{{\ul 0.830}} &
  0.135 \\ \hline
\textbf{Leiden-Mod} &
  \multicolumn{1}{l|}{0.220} &
  \multicolumn{1}{l|}{0.271} &
  0.221 &
  \multicolumn{1}{l|}{0.022} &
  \multicolumn{1}{l|}{0.073} &
  0.132 &
  \multicolumn{1}{l|}{0.493} &
  \multicolumn{1}{l|}{0.503} &
  0.153 \\ \hline
\textbf{Infomap} &
  \multicolumn{1}{l|}{0.167} &
  \multicolumn{1}{l|}{0.233} &
  0.245 &
  \multicolumn{1}{l|}{0.010} &
  \multicolumn{1}{l|}{0.081} &
  0.163 &
  \multicolumn{1}{l|}{0.490} &
  \multicolumn{1}{l|}{0.454} &
  0.303 \\ \hline
\end{tabular}%
}
\end{table}


We then investigated the accuracy of methods after post-processing by WCC. Since the synthetic network ground truth clusters are all well-connected, WCC is expected to be beneficial.
As seen in Supplementary Materials Figure S2, the most accurate method is Leiden-CPM(0.01)+WCC, achieving the same top accuracy for AMI and NMI  as Leiden-Mod+WCC, but achieving better ARI scores than Leiden-Mod+WCC.
Based on this, we select Leiden-CPM(0.01)+WCC as our choice for use as the final clustering step in our ensemble approach.


\subsection{Experiment 3: Designing the cluster ensemble method}

The  cluster ensemble method we design takes as input a network, a set of two or more clusterings, a threshold $t$, and a technique for clustering the final edge-weighted network.
Based on Supplementary Materials Fig S2,  we selected Leiden-CPM(0.01)+WCC as the clustering method for the final edge-weighted network.
Here we describe the rest of the design.

\vspace{-.1in}
\paragraph{Selecting the clustering methods to combine}

Since we are using Leiden-CPM for the final clustering method, we do not consider Leiden-CPM for the input clusterings. 
We sought  clustering algorithms with complementary characteristics in terms of precision, recall, and FPR (see Supplementary Material Figure S3).
DSC-Flow-Iter exhibits moderate precision and a low FPR, but also suffers from low recall. 
This indicates that while the pairs of nodes DSC-Flow-Iter places in the same community are very likely correct, it misses many other pairs that should be together. Conversely, Leiden-Mod shows a very high recall but has low precision and a high FPR, which   suggests that modularity optimization tends to group nodes aggressively, correctly co-clustering many pairs but also incorrectly merging distinct communities, which is consistent with theoretical observations about the resolution limit made in \cite{fortunato2007resolution}. 
IKC shows a profile similar to DSC-Flow-Iter but is outperformed on all metrics. Infomap is similar to Leiden-Mod but is also outperformed on all metrics, especially with respect to FPR. 
The complementary properties of  DSC-Flow-Iter (high precision, low recall) and Leiden-Mod (low precision, high recall) make them good choices for an ensemble, and we use this combination in the subsequent experiment.

\vspace{-.1in}
\paragraph{Comparing different cluster ensemble variants}

Based on the previous experiment, we explored the cluster ensemble approach using 
 DSC-Flow-Iter and Leiden-Mod as the input clusterings and
 Leiden-CPM(0.01)+WCC to cluster the weighted network we produce.
 What remains to decide is the value of the threshold $t$ and whether Leiden should consider the edge weights. 
 We consider 
  two   values  for $t$ ($0.5$ and $1.0$) and vary whether we consider edge weights, thus producing four variants of the cluster ensemble.

We summarize the trends we observed (see  Supplementary Materials Figure S4 for full results).
Setting $t=0.5$ improved accuracy for all three criteria, but
considering weights was slightly detrimental for accuracy.
Thus, using the unweighted network generated from the combination with a threshold of $t=0.5$ provides the best and most consistent gains. 
Furthermore, this variant improved accuracy in over 75\% of the networks tested.
Thus, we propose:
\begin{enumerate}
\item Stage 1: Run DSC-Flow-Iter on the input network
\item Stage 2: Run Leiden-Mod on the input network
\item Stage 3: Combine these clusterings to create an unweighted output network, retaining only the edges with weights at least threshold $t = 0.5$
\item Stage 4: Run Leiden-CPM(0.01) on this output network and apply WCC post-processing to the resulting clustering
\end{enumerate}

\begin{table}[]
\centering
\setlength{\tabcolsep}{4pt}
\renewcommand{\arraystretch}{1.5}
\caption[]{\textbf{Accuracy of the ensemble and all its constituents} The results are on $74$ synthetic EC-SBM networks with SBM+WCC ground-truth clustering.  Leiden-CPM+WCC is run with resolution value 0.01. The ensemble method applies Leiden-CPM+WCC to the graph formed on Flow-Iter and Leiden-Mod, using threshold 0.5.
The best accuracy is boldfaced.}
\label{table:ensemble-accuracy}
\resizebox{\textwidth}{!}{%
\begin{tabular}{|l|rrr|rrr|rrr|}
\hline
\multicolumn{1}{|c|}{\multirow{2}{*}{}} &
  \multicolumn{3}{c|}{\textbf{AMI}} &
  \multicolumn{3}{c|}{\textbf{ARI}} &
  \multicolumn{3}{c|}{\textbf{NMI}} \\ \cline{2-10} 
\multicolumn{1}{|c|}{} &
  \multicolumn{1}{c|}{\textbf{Med}} &
  \multicolumn{1}{c|}{\textbf{Avg}} &
  \multicolumn{1}{c|}{\textbf{Std}} &
  \multicolumn{1}{c|}{\textbf{Med}} &
  \multicolumn{1}{c|}{\textbf{Avg}} &
  \multicolumn{1}{c|}{\textbf{Std}} &
  \multicolumn{1}{c|}{\textbf{Med}} &
  \multicolumn{1}{c|}{\textbf{Avg}} &
  \multicolumn{1}{c|}{\textbf{Std}} \\ \hline
\textbf{DSC-Flow-Iter} &
  \multicolumn{1}{r|}{0.352} &
  \multicolumn{1}{r|}{0.370} &
  0.211 &
  \multicolumn{1}{r|}{0.053} &
  \multicolumn{1}{r|}{0.121} &
  0.157 &
  \multicolumn{1}{r|}{0.876} &
  \multicolumn{1}{r|}{0.833} &
  0.129 \\ \hline
\textbf{Leiden-Mod} &
  \multicolumn{1}{r|}{0.220} &
  \multicolumn{1}{r|}{0.271} &
  0.221 &
  \multicolumn{1}{r|}{0.022} &
  \multicolumn{1}{r|}{0.073} &
  0.132 &
  \multicolumn{1}{r|}{0.493} &
  \multicolumn{1}{r|}{0.503} &
  0.153 \\ \hline
\textbf{Leiden-CPM+WCC} &
  \multicolumn{1}{r|}{0.515} &
  \multicolumn{1}{r|}{0.496} &
  0.288 &
  \multicolumn{1}{r|}{0.172} &
  \multicolumn{1}{r|}{0.271} &
  0.273 &
  \multicolumn{1}{r|}{0.914} &
  \multicolumn{1}{r|}{0.882} &
  0.141 \\ \hline
\textbf{Ensemble} &
  \multicolumn{1}{r|}{\bf{0.572}} &
  \multicolumn{1}{r|}{\bf{0.516}} &
  0.301 &
  \multicolumn{1}{r|}{\bf{0.225}} &
  \multicolumn{1}{r|}{\bf{0.308}} &
  0.281 &
  \multicolumn{1}{r|}{\bf{0.921}} &
  \multicolumn{1}{r|}{\bf{0.894}} &
  0.136 \\ \hline
\end{tabular}%
}
\end{table}
\vspace{-.1in}
\paragraph{Results for the ensemble method}

As seen in Table \ref{table:ensemble-accuracy}, using our proposed ensemble method achieved better accuracy than its constituent methods, for all three criteria (AMI, ARI, and NMI). Specifically, running Leiden-CPM(0.01)+WCC on the modified network based on clustering of  DSC-Flow-Iter and Leiden-Mod gave a more accurate  clustering than running on the original network. Furthermore, as we have previously identified Leiden-CPM(0.01)+WCC as the most accurate single-algorithm method, this indicates that the selected ensemble method provides an improvement over all the single-algorithm techniques.



\subsection{Experiment 4: Runtime for the pipeline}

We evaluated the runtime of our proposed pipeline on 78 real-world networks. The pipeline failed to complete on five large networks (\texttt{google\_web}, \texttt{livejournal}, \texttt{orkut}, \texttt{bitcoin}, and \texttt{CEN}) ranging from $\sim 850K$ to $\sim 14M$ nodes, due to OOM (out-of-memory) errors during Stage 1. 

\begin{figure}[!ht]
    \centering
    \includegraphics[width=\linewidth]{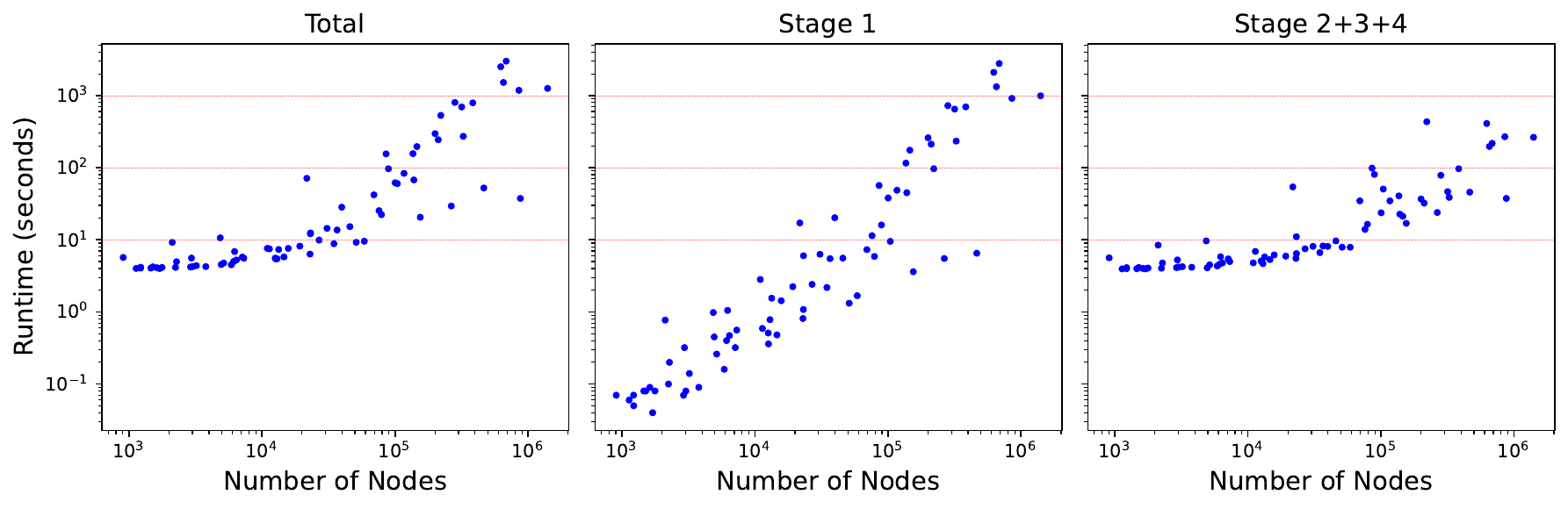}
    \caption[Scatter plot of runtimes of the pipeline on real-world networks]{\textbf{Scatter plot of runtimes of the pipeline on real-world networks} Both $x$-axis and $y$-axis are in log-scale. The results are on $73$ real-world networks. Stage 1 runs DSC-Flow-Iter. 
    We observe that the pipeline tends to run longer as network size grows. Furthermore, the dominant factor is Stage 1 (DSC-Flow-Iter) for larger networks but not for smaller networks.}
    \label{fig:exp4_scatter}
\end{figure}

\begin{figure}[!ht]
    \centering
    \includegraphics[width=\linewidth]{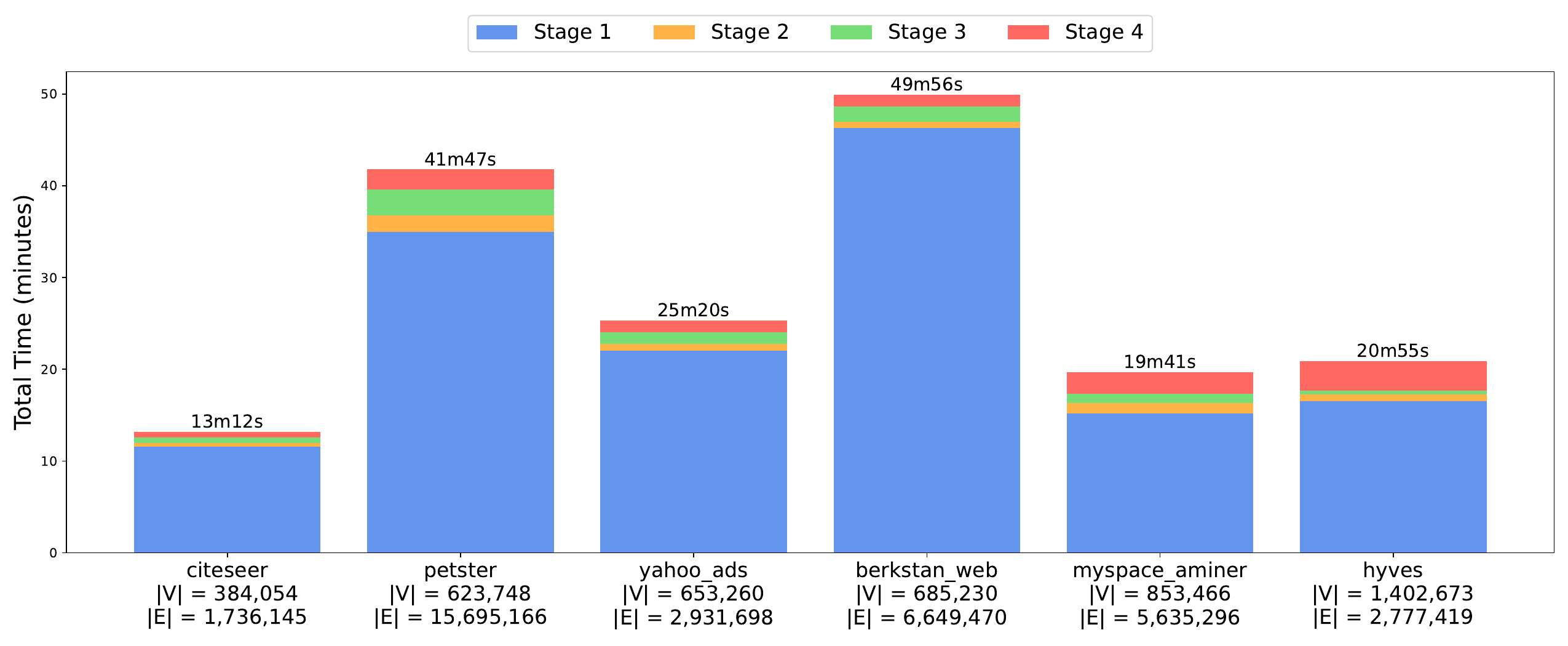}
    \caption[Breakdown of pipeline runtime by stage on those networks that used at least 800 seconds]{\textbf{Breakdown of pipeline runtime by stage  on networks that used at least 800 seconds} Stage 1 runs DSC-Flow-Iter. Stage 2 runs Leiden-Mod. Stage 3 uses the clusterings from the previous two stages to generate the unweighted network. Stage 4 applies the final clustering (Leiden-CPM(0.01)+WCC)).}
    \label{fig:exp4}
\end{figure}

As shown in Figure \ref{fig:exp4_scatter}, the runtime of Stage 1 exhibits a power-law relationship with the network size, appearing as a linear trend on the log-log scale, but the combined runtime of the remaining stages grows more slowly. This difference in growth rates means that while the rest of the pipeline (Stages 2, 3, and 4) can dominate the runtime for smaller networks, Stage 1 becomes the clear bottleneck and the most time-consuming step for larger networks. This is also evident from the breakdown of the runtime in Figure \ref{fig:exp4}. These trends highlight that the scalability of our current pipeline is limited by the DSC-Flow-Iter stage.

\section{Conclusion}
\label{sec:conclusions}

In this work, we have introduced a novel community detection algorithm, DSC-Flow-Iter, based on the iterative extraction of dense subgraphs. Our experiments showed that this method achieved high accuracy,  second only to Leiden-CPM(0.01), on a suite of synthetic networks. More importantly, we identified the complementary nature of our density-based method (high precision) and modularity-based methods like Leiden-Mod (high recall). 
This insight led us to propose a new cluster ensemble method that integrates these different approaches. 
The resulting pipeline, which combines the outputs of DSC-Flow-Iter and Leiden-Mod to generate a refined network, followed by a final clustering with Leiden-CPM(0.01) and post-processing with WCC, improved on the best single-algorithm approach. Overall, therefore, we find that while community detection via dense subgraphs is not the best on its own, it can contribute within  an ensemble of diverse methods, which together achieve a better accuracy.

The ensemble technique can also be described as a way of using the input clusterings (here, Leiden-Mod and DSC-Flow-Iter) to identify the unreliable edges in the network, so that they can be removed before the network is clustered (here by Leiden-CPM(0.01)+WCC).  The entire improvement in accuracy over Leiden-CPM(0.01)+WCC, therefore, is due to the identification and deletion of these unreliable edges. 
We also note that the specific methods used as the input clusterings {\em does matter}, as replacing Leiden-Mod by Infomap (Supplementary Materials Figure S5) did {\em not} produce as large an improvement in accuracy over Leiden-CPM(0.01)+WCC as using Leiden-mod.

Thus, future work 
 should explore other ways of designing cluster ensemble methods, both for the selection of the input clusterings and in how they are used. 
 Future work also needs to explore a larger set of synthetic  networks to better understand the specific network properties  where this and other cluster ensemble techniques provide benefits.  
Since the pipeline  has a high memory requirement on some large networks, due to the DSC-Flow-Iter step,  a new implementation that addresses memory usage is needed.

%
%


\vspace{-.1in}
\paragraph{Funding}
This work was supported in part by the Illinois-Insper partnership and by NSF grant 2402559 to TW and GC.
\bibliographystyle{spmpsci} 
\bibliography{clustering} 
\end{document}


%
\title{Supplementary Materials to ``Dense Subgraph Clustering and a new Cluster Ensemble method"}
%
%
\author{The-Anh Vu-Le \and Minhyuk Park  \and Ian Chen \and Jo\~{a}o Alfredo Cardoso Lamy  \and Tom\'as Alessi  \and Elfarouk Harb  \and 
George Chacko \and Tandy Warnow
}
%
%


\maketitle 

\tableofcontents
\listoffigures
\listoftables

\clearpage

\section{Dataset}

\paragraph{Networks (76) from \cite{netzschleuder} sorted by increasing node count}

\texttt{dnc}, \texttt{uni\_email}, \texttt{polblogs}, \texttt{faa\_routes}, \texttt{netscience}, \texttt{new\_zealand\_collab}, \texttt{collins\_yeast}, 

\noindent \texttt{interactome\_stelzl}, \texttt{bible\_nouns}, \texttt{at\_migrations}, \texttt{interactome\_figeys}, 

\noindent \texttt{us\_air\_traffic}, \texttt{drosophila\_flybi}, \texttt{fly\_larva}, \texttt{interactome\_vidal}, 

\noindent \texttt{openflights}, \texttt{bitcoin\_alpha}, \texttt{fediverse}, \texttt{power}, \texttt{advogato}, \texttt{bitcoin\_trust}, \texttt{jung}, \texttt{reactome}, \texttt{jdk}, \texttt{elec}, \texttt{chess}, \texttt{sp\_infectious}, \texttt{wiki\_rfa}, \texttt{dblp\_cite}, \texttt{anybeat}, \texttt{chicago\_road}, \texttt{foldoc}, \texttt{inploid}, \texttt{google}, \texttt{marvel\_universe}, \texttt{fly\_hemibrain}, \texttt{internet\_as}, \texttt{word\_assoc}, \texttt{cora}, \texttt{lkml\_reply}, \texttt{linux}, \texttt{topology}, \texttt{email\_enron}, \texttt{pgp\_strong}, \texttt{facebook\_wall}, \texttt{slashdot\_threads}, \texttt{python\_dependency}, \texttt{marker\_cafe}, 

\noindent \texttt{epinions\_trust}, \texttt{slashdot\_zoo}, \texttt{twitter\_15m}, \texttt{prosper}, \texttt{wiki\_link\_dyn}, 

\noindent \texttt{livemocha}, \texttt{wikiconflict}, \texttt{lastfm\_aminer}, \texttt{wiki\_users}, \texttt{wordnet}, \texttt{douban}, 

\noindent \texttt{academia\_edu}, \texttt{google\_plus}, \texttt{libimseti}, \texttt{email\_eu}, \texttt{stanford\_web}, 

\noindent \texttt{dblp\_coauthor\_snap}, \texttt{notre\_dame\_web}, \texttt{citeseer}, \texttt{twitter}, \texttt{petster}, \texttt{yahoo\_ads}, \texttt{berkstan\_web}, \texttt{myspace\_aminer}, \texttt{google\_web}, \texttt{ hyves}, \texttt{livejournal}, \texttt{bitcoin}

\paragraph{Other networks (2) from \cite{cm-journal} sorted by increasing node count}

\texttt{orkut}, \texttt{CEN} (Curated Exosome Network)

\clearpage

\section{Software and Commands}

\subsection{Leiden}

We use the \texttt{leidenalg} Python package for Leiden. A wrapper script is in \cite{dsc-github} at \texttt{src/leiden/run\_leiden.py}.

\paragraph{Leiden-CPM} To run Leiden optimizing under the Constant Potts Model, the command is
\begin{lstlisting}
python src/leiden/run_leiden.py \
    --edgelist <edgelist> \
    --output-directory <output_dir> \
    --model cpm \
    --resolution <resolution>
\end{lstlisting}
where
\begin{itemize}
    \item \texttt{<edgelist>}: path to the input edgelist file (TSV format)
    \item \texttt{<output\_dir>}: path to the output directory where the results will be saved
    \item \texttt{<resolution>}: resolution parameter for the Constant Potts Model (recommended: 0.01)
\end{itemize}

\paragraph{Leiden-Mod} To run Leiden optimizing modularity, the command is
\begin{lstlisting}
python src/leiden/run_leiden.py \
    --edgelist <edgelist> \
    --output-directory <output_dir> \
    --model mod
\end{lstlisting}
where \texttt{<edgelist>} and \texttt{<output\_dir>} are the same as in the previous command.

\subsection{Infomap}

We use the \texttt{infomap} Python package for Infomap. A wrapper script is in \cite{dsc-github} at \texttt{src/infomap/run\_infomap.py}.

To run Infomap, the command is
\begin{lstlisting}
python src/infomap/run_infomap.py \
    --edgelist <edgelist> \
    --output-directory <output_dir>
\end{lstlisting}
where
\begin{itemize}
    \item \texttt{<edgelist>}: path to the input edgelist file (TSV format)
    \item \texttt{<output\_dir>}: path to the output directory where the results will be saved
\end{itemize}

\subsection{IKC}

We use the script in \cite{dsc-github} at \texttt{src/ikc/run\_ikc.py}.

To run IKC, the command is
\begin{lstlisting}
python src/ikc/run_ikc.py \
    --edgelist <edgelist> \
    --output-directory <output_dir> \
    --kvalue <k>
\end{lstlisting}
where
\begin{itemize}
    \item \texttt{<edgelist>}: path to the input edgelist file (TSV format)
    \item \texttt{<output\_dir>}: path to the output directory where the results will be saved
    \item \texttt{<k>}: the minimum value of $k$ allowed for a $k$-core to be included (recommended: 1)
\end{itemize}

\subsection{Density-based methods}

We use the binaries from the DSC repository \cite{dsc-github}.

\paragraph{DSC-Flow-Iter} To run DSC-Flow-Iter, the command is
\begin{lstlisting}
./bin/flow-iter <edgelist> <com> <density>
\end{lstlisting}
where
\begin{itemize}
    \item \texttt{<edgelist>}: path to the input edgelist file (TSV format)
    \item \texttt{<com>}: path to the output community file (TSV format)
    \item \texttt{<density>}: path to the output density file (TSV format)
\end{itemize}

\paragraph{DSC-Flow} To run DSC-Flow, replace \texttt{./bin/flow-iter} with \texttt{./bin/flow} in the above command.

\paragraph{DSC-FISTA(int)} To run DSC-FISTA(int), the command is
\begin{lstlisting}
./bin/fista-int <niters> <edgelist> <com> <density>
\end{lstlisting}
where
\begin{itemize}
    \item \texttt{<niters>}: number of iterations to run (recommended: 200)
    \item \texttt{<edgelist>}: path to the input edgelist file (TSV format)
    \item \texttt{<com>}: path to the output community file (TSV format)
    \item \texttt{<density>}: path to the output density file (TSV format)
\end{itemize}

\paragraph{DSC-FISTA-Iter} To run DSC-FISTA-Iter, replace \texttt{./bin/fista-int} with \texttt{./bin/fista-frac} in the above command.

\subsection{Ensemble}

We use the binary in \cite{clustermerger-github}.

To run the ensemble clustering merger, the command is
\begin{lstlisting}
./cluster_merger Weighted \
    --edgelist <edgelist> \
    --clustering-list <clustering_list> \
    --weighting-strategy 0 \
    --threshold <threshold> \
    --output-file "" \
    --output-weighted-graph <output_fn> \
    --log-file <log_file>
\end{lstlisting}
where
\begin{itemize}
    \item \texttt{<edgelist>}: path to the input edgelist file (TSV format with source and target integer node IDs on each row)
    \item \texttt{<clustering\_list>}: path to a file with a list of clustering file paths (one per line)
    \item \texttt{<threshold>}: threshold for edge weights; edges with weights below this value are removed (default: -1, recommended: 0.5)
    \item \texttt{<output\_fn>}: path to the output weighted graph file (TSV format with columns: source, target, weight)
    \item \texttt{<log\_file>}: path to the log file
\end{itemize}

\subsection{Pipeline}

We use the script \texttt{pipeline.sh} in \cite{dsc-github} for running the recommended pipeline.

To run the pipeline, the command is
\begin{lstlisting}
bash pipeline.sh <edgelist> <output_directory>
\end{lstlisting}
where
\begin{itemize}
    \item \texttt{<edgelist>}: path to the input edgelist file (TSV format)
    \item \texttt{<output\_directory>}: path to the output directory where the results will be saved
\end{itemize}

\clearpage

\section{Additional Tables}

\begin{table}[!ht]
\centering
\setlength{\tabcolsep}{4pt}
\renewcommand{\arraystretch}{1.5}
\caption[Node coverage of different methods]{\textbf{Node coverage of different methods} The results are on $74$ synthetic EC-SBM networks with SBM+WCC ground-truth clustering.}
\begin{tabular}{|l|r|r|r|}
\hline
\textbf{} & \multicolumn{1}{c|}{\textbf{Median}} & \multicolumn{1}{c|}{\textbf{Avg}} & \multicolumn{1}{c|}{\textbf{Std}} \\ \hline
\textbf{DSC-Flow-Iter}    & 0.659328 & 0.635079 & 0.194500 \\ \hline
\textbf{Leiden-Mod}       & 1.000000 & 0.999999 & 0.000006 \\ \hline
\textbf{Leiden-CPM(0.01)} & 0.891256 & 0.837698 & 0.163163 \\ \hline
\textbf{Infomap}          & 1.000000 & 0.999992 & 0.000037 \\ \hline
\textbf{IKC(1)}           & 0.628032 & 0.607151 & 0.198815 \\ \hline
\end{tabular}
\end{table}

\clearpage

\section{Additional Figures}

\begin{figure}[!ht]
    \centering
    \includegraphics[width=\linewidth]{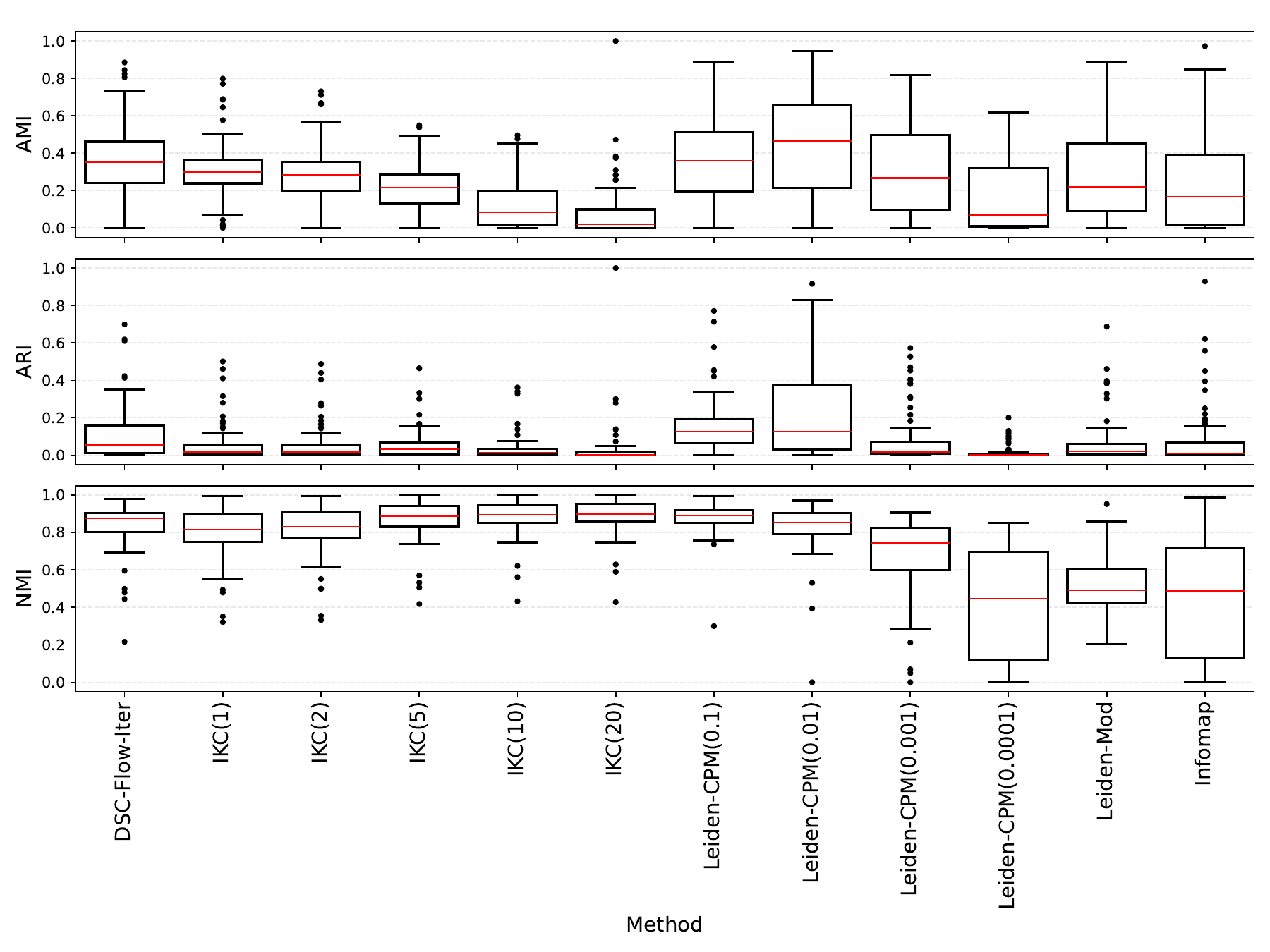}
    \caption[Accuracy of DSC-Flow-Iter in comparison to other 11 clustering methods]{\textbf{Accuracy of DSC-Flow-Iter in comparison to 11 clustering methods} Results are shown for $74$ synthetic EC-SBM networks with SBM+WCC ground-truth clustering.}
    \label{fig:suppl_exp2}
\end{figure}

\begin{figure}[!ht]
    \centering
    \includegraphics[width=\linewidth]{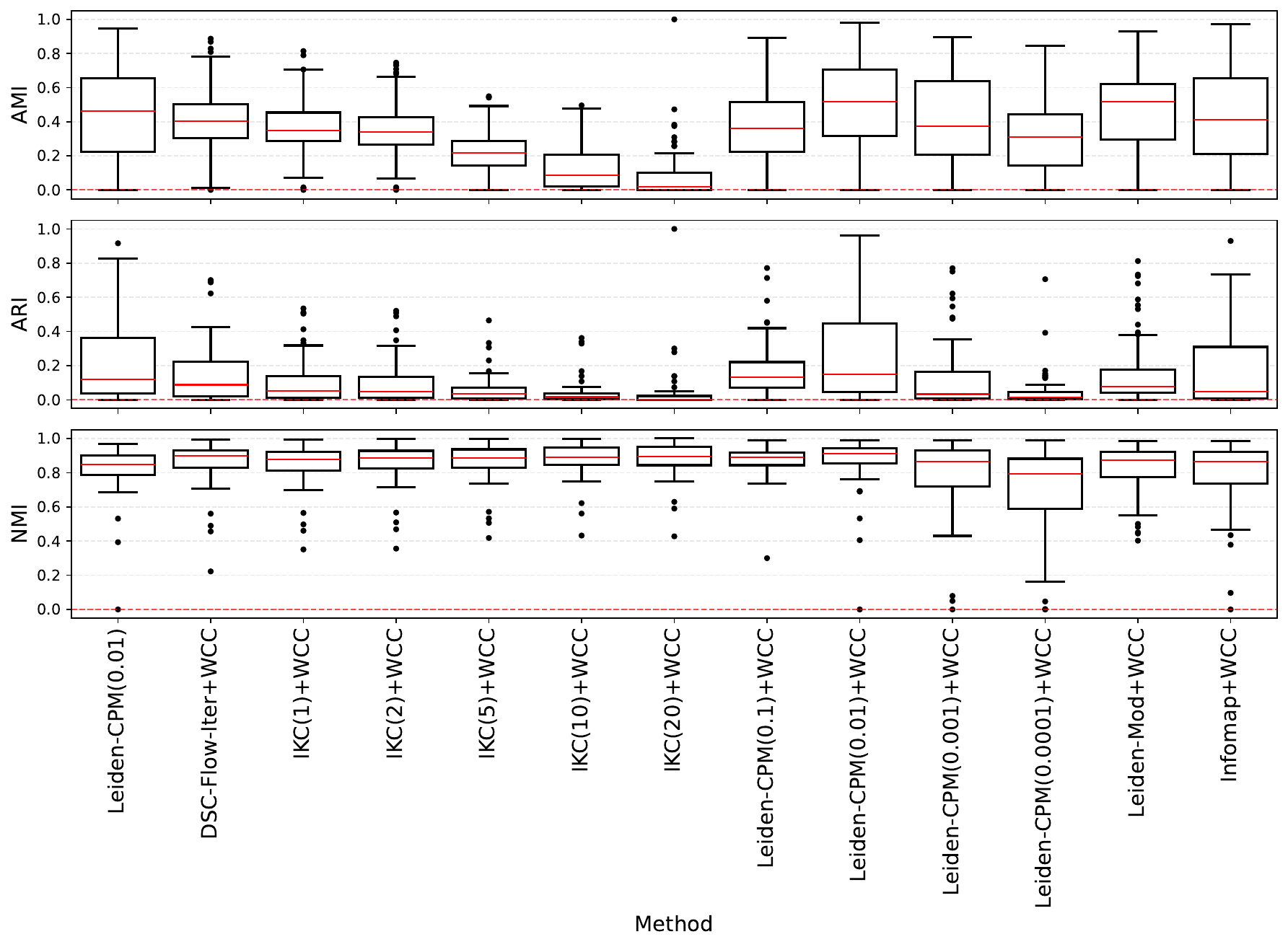}
    \caption[Accuracy of 12 community detection methods, 11 with WCC treatment]{\textbf{Accuracy of 12 community detection methods, 11 with  WCC treatment} The results are on $66$ synthetic EC-SBM networks with SBM+WCC ground-truth clustering (the WCC post-processing failed on 8 of the 74 networks, due to OOM errors). This shows that Leiden-CPM(0.01)+WCC was the best of the studied methods.}
    \label{fig:suppl_exp2_wcc}
\end{figure}

\begin{figure}[!ht]
    \centering
    \includegraphics[width=\linewidth]{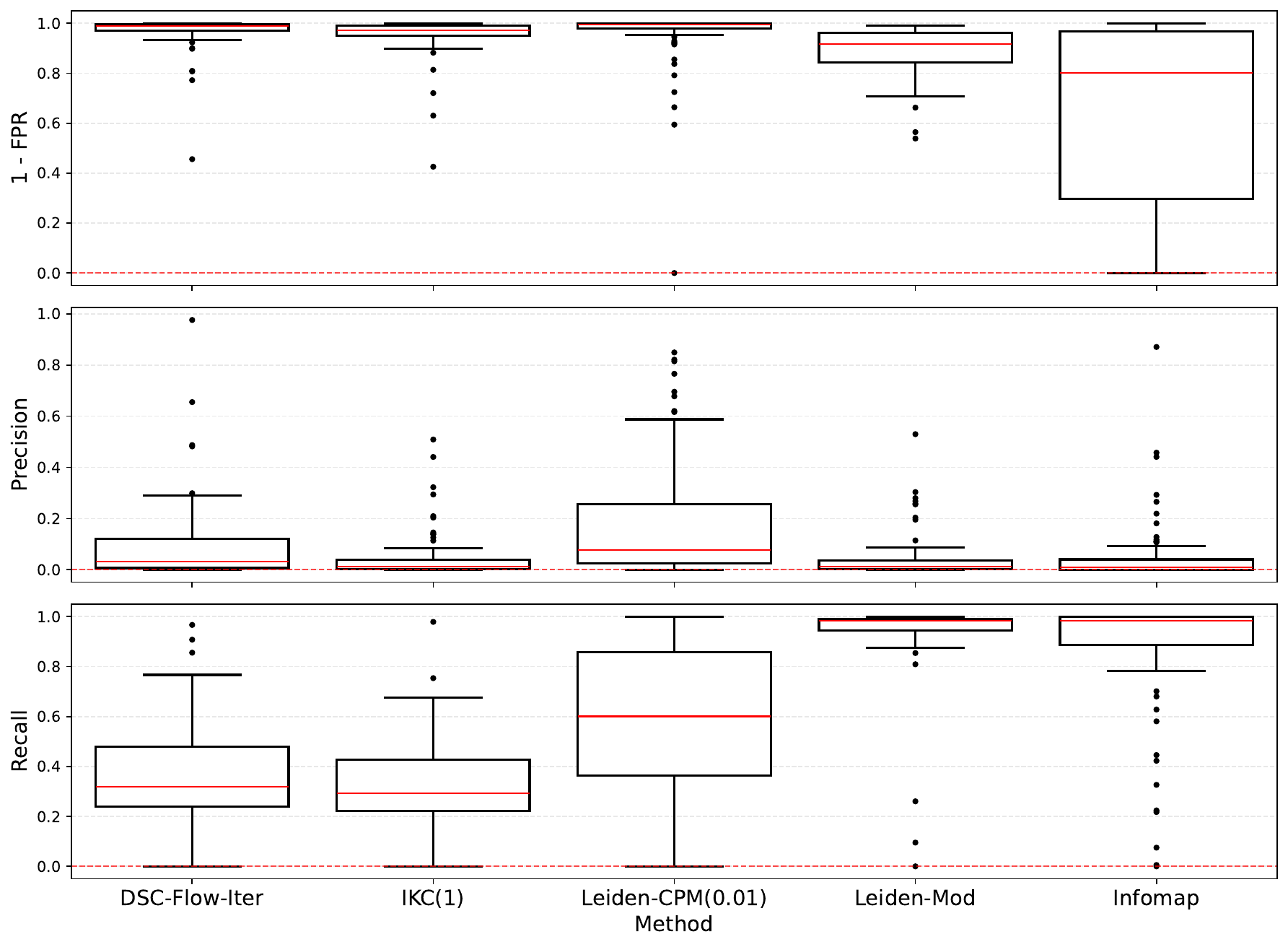}
    \caption[Precision, recall, and 1-FPR of five clustering methods]{\textbf{Precision, recall, and 1-FPR of five clustering methods} The results are on $74$ synthetic EC-SBM networks with SBM+WCC ground-truth clustering.}
    \label{fig:exp3_confusion-selected}
\end{figure}


\begin{figure}[!ht]
    \centering
    \includegraphics[width=\linewidth]{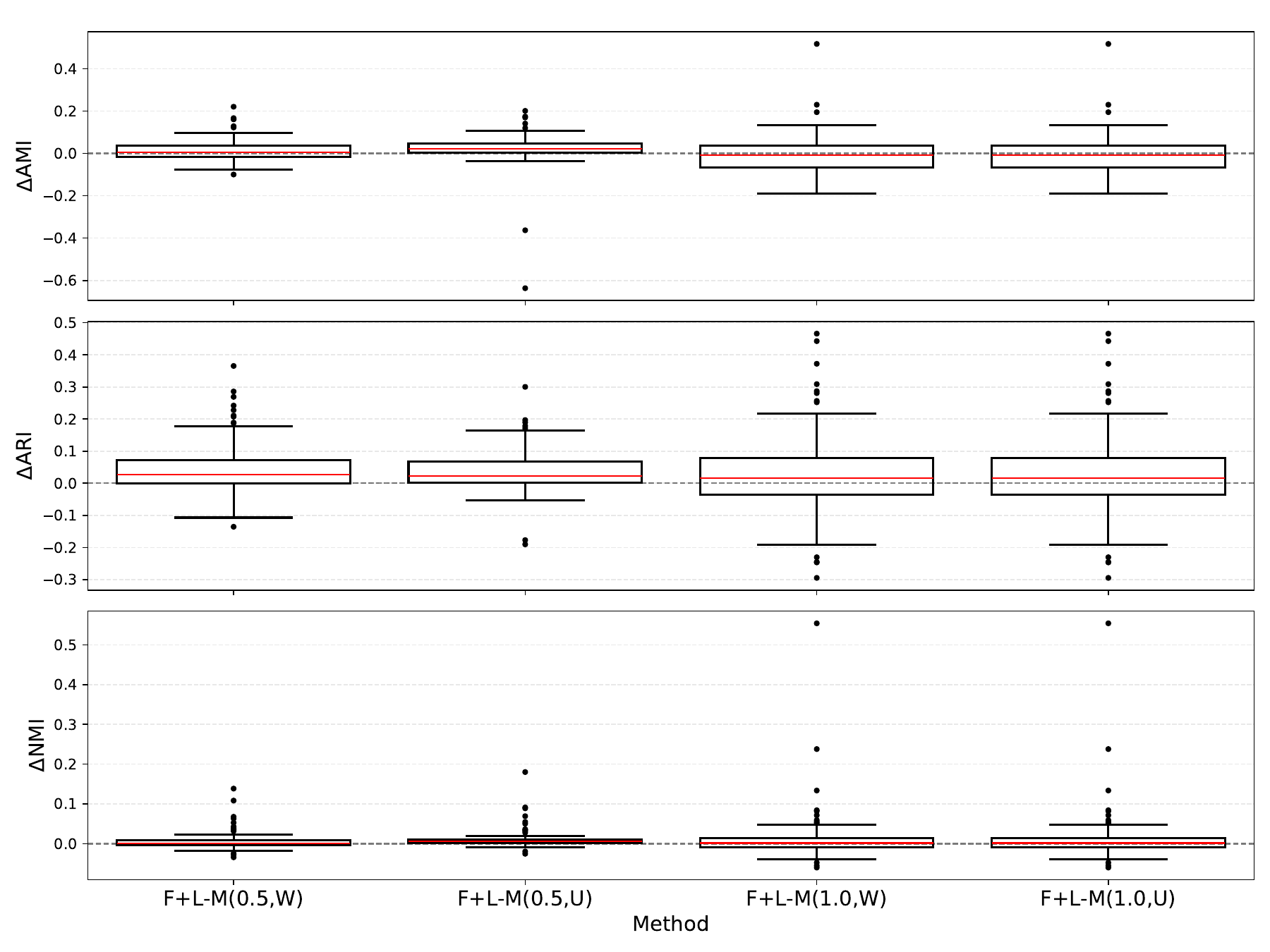}
    \caption[Difference in accuracy of 4 variants of the Ensemble method combining Flow-Iter and Leiden-Mod clusterings relative to Leiden-CPM(0.01)+WCC]{\textbf{Difference in accuracy of 4 variants of the Ensemble method combining Flow-Iter and Leiden-Mod clusterings relative to Leiden-CPM(0.01)+WCC} The variants differ in the threshold $t$ and whether the network edge weights are considered. A positive difference means the method has higher accuracy than Leiden-CPM(0.01)+WCC, and vice versa. The results are on $74$ synthetic EC-SBM networks with SBM+WCC ground-truth clustering. F stands for DSC-Flow-Iter. L stands for Leiden-Mod. M stands for Merged. The  first value in the parenthesis after M is the threshold $t$, while W and U respectively stand for Weighted and Unweighted. All ensembles are processed with WCC (i.e., they are Leiden-CPM(0.01)+WCC results).  All variants of the Ensemble technique with $t=0.5$ improve on Leiden-CPM(0.01)+WCC, but the variants with $t=1.0$   have lower AMI scores.  The best results are obtained using  DSC-Flow-Iter with Leiden-Mod, setting $t=0.5$ and not considering weights (i.e., F+L-M(0.5, U)), which is helpful for 75\% of the cases.
    }
    \label{fig:exp3_leiden_fl_u_diff-selected}
\end{figure}

\begin{figure}[!ht]
    \centering
    \includegraphics[width=\linewidth]{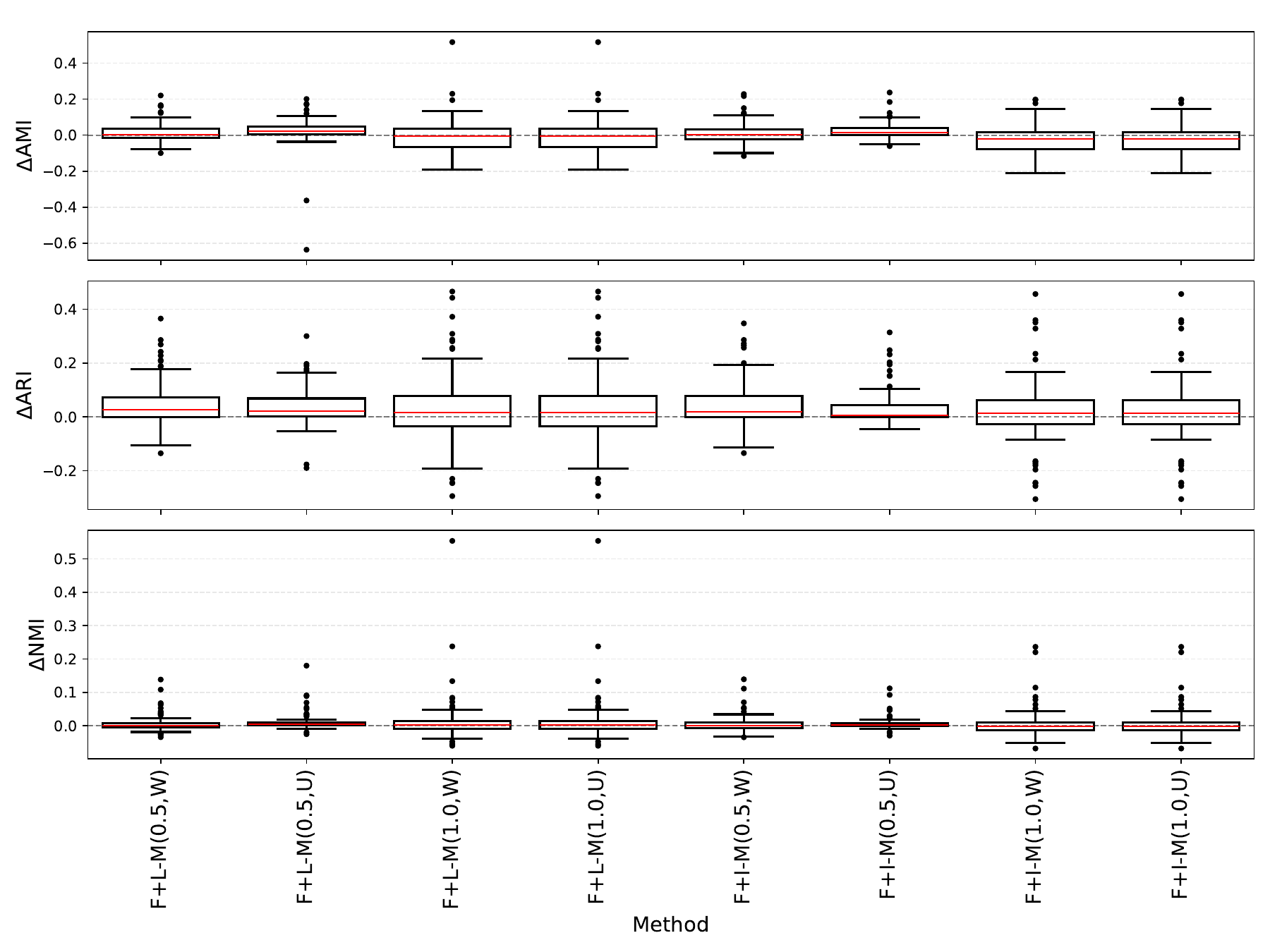}
    \caption[Comparison of 8 variants of the Ensemble method]{\textbf{Difference in accuracy of 8 variants of the Ensemble method of clusterings relative to Leiden-CPM(0.01)+WCC} The variants combine Flow-Iter (F) with either Leiden-mod (L) or Infomap (I), and differ by threshold $t$ (which is set to $0.5$ or $1.0$) and whether the network weights are considered.   A positive difference means the method has higher accuracy than Leiden-CPM(0.01)+WCC, and vice versa. The results are on $74$ synthetic EC-SBM networks with SBM+WCC ground-truth clustering. M stands for Merged; the value in the parentheses after M  are the threshold $t$ and whether the network is weighted (W) or unweighted (U). All ensembles are processed with WCC (i.e., they are Leiden-CPM(0.01)+WCC results). All variants of the Ensemble technique with $t=0.5$ improve on Leiden-CPM(0.01)+WCC, but the variants with $t=1.0$  have lower AMI scores. Results with Infomap instead of Leiden-Mod have lower accuracy. The best results are obtained using DSC-Flow-Iter with Leiden-Mod, setting $t=0.5$ and not considering weights (i.e., F+L-M(0.5, U)), which is helpful for around 75\% of the cases.}
    \label{fig:exp3_leiden_fl_u_diff-all}
\end{figure}



\clearpage

\bibliographystyle{spmpsci}
\bibliography{clustering}